\documentclass[useAMS,usenatbib]{mn2e}
\usepackage{latexsym,graphicx,natbib}
\usepackage{amsmath,subfigure}

\def\fun#1#2{\lower3.6pt\vbox{\baselineskip0pt\lineskip.9pt
  \ialign{$\mathsurround=0pt#1\hfil##\hfil$\crcr#2\crcr\sim\crcr}}}
\def\lap{\mathrel{\mathpalette\fun <}}
\def\gap{\mathrel{\mathpalette\fun >}}

\def\mbh{{M_{\rm BH}}}
\def\msun{M_{\odot}}
\def\mpc{\,\rm mpc}
\def\pc{\,\rm pc}
\def\kms{{\rm\,km\,s^{-1}}} 
\def\beq{\begin{equation}}
\def\eeq{\end{equation}}

\newcommand{\mh}{M_\bullet}

\newcommand{\myddot}[1]{\mathbf{#1}^{(2)}}
\newcommand{\mydddot}[1]{\mathbf{#1}^{(3)}}

\title[A Hybrid $N$-body code]{A Hybrid $N$-Body Code Incorporating Algorithmic Regularization
and Post-Newtonian Forces}

\author[Harfst et al.]{S.~Harfst,$^1$\thanks{harfst@science.uva.nl} A.~Gualandris,$^2$ D.~Merritt$^2$ and S.~Mikkola$^2$\\
$^1$Sterrenkundig Instituut ``Anton Pannekoek'' and Section Computational Science, 
University of Amsterdam,\\ Kruislaan 403, 1098SJ Amsterdam, The Netherlands\\
$^2$Center for Computational Relativity and Gravitation, 
Rochester Institute of Technology, 
78 Lomb Memorial Drive, Rochester, NY 14623\\
$^3$Tuorla Observatory, University of Turku, V\"ais\"al\"antie 20, Piikki\"o, Finland
}

\begin{document}

\maketitle

\begin{abstract}
We describe a novel $N$-body code designed for simulations of 
the central regions of galaxies containing massive black holes.
The code incorporates Mikkola's ``algorithmic'' chain 
regularization scheme including post-Newtonian terms up to 
PN2.5 order.
Stars moving beyond the chain are advanced using a fourth-order 
integrator with forces computed on a GRAPE board.
Performance tests confirm that the hybrid code achieves 
better energy conservation, in less elapsed time, than the 
standard scheme and that it reproduces the orbits of
stars tightly bound to the black hole with high precision.
The hybrid code is applied to two sample problems:
the effect of finite-$N$ gravitational fluctuations on the
orbits of the S-stars;
and inspiral of an intermediate-mass black hole into the
galactic center. 
\end{abstract}


\section{Introduction}

Standard $N$-body integrators have difficulty reproducing the motion
of tight binaries, or following close hyperbolic encounters between
stars.  Unless the integration time step is made very short, the
relative orbit will not be accurately reproduced, and the system
energy may also exhibit an unacceptable drift since a single compact
subsystem can contain a large fraction of the total binding energy.
This difficulty is particularly limiting for studies of the centers of
galaxies like the Milky Way, where the gravitational potential is
dominated by a (single or binary) supermassive black hole.  Early
attempts to simulate such systems often had difficulty achieving the
required accuracy and performance \citep[e.g.][]{SHQ:95}.  In later
studies, stars in tightly bound orbits around the black hole were
sometimes simply removed \citep[e.g.][]{BME:04,MME:07}; in other
studies, the position and velocity of the most massive particle were
artificially fixed and the stellar orbits approximated as
perturbed Keplerian ellipses \citep[e.g.][]{LB:08}.  While these
approaches can be useful for certain problems, a less restrictive
treatment is indicated for studies in which the motion of stars near
the black hole is used to constrain the magnitude of non-Keplerian
perturbations \citep{fragile-00,weinberg-05,will-08}.  Fixing the
location of the most massive particle can also be problematic when
simulating binary or multiple supermassive black holes.

\citet{aarseth-03a} summarizes the various regularized schemes that
have been incorporated into N-body codes to treat strong gravitational
interactions with high accuracy and without loss of performance.

\noindent
(1) The KS-regularization method \citep{KS-65}, together with the
original chain method \citep{MA-93} which uses the KS-transformation
to regularize multiparticle systems, has been widely used to integrate
binaries in simulations of star clusters.  This method has also been
applied to the binary black hole problem
\citep[e.g.][]{QH-97,MM-01,MMS:07}.  This approach suffers from
inaccuracies when mass ratios are very large because the total energy
that appears in the equations of motion is dominated by the binary
\citep{aarseth-03b}.

\noindent
(2) In the case of a single dominant body, an alternative to the 
chain geometry, called wheel-spoke (WS) regularization, treats 
each interaction with the massive body via the standard KS-regularization,
while other interactions use a small softening to avoid singularities 
\citep{zare-74,aarseth-07}.

\noindent
(3) Algorithmic regularization (AR) methods effectively remove 
singularities with a time transformation accompanied by the leapfrog 
algorithm. 
There are two such methods: the logarithmic Hamiltonian (LogH) 
\citep{MT-99a,MT-99b,preto-99} and time-transformed leapfrog (TTL) 
\citep{MA-02}. 
Both these algorithms produce exact trajectories for the unperturbed 
two-body problem and provide regular results for more general cases. 
In the LogH method the derivative of time is inversely proportional to 
the gravitational potential while in TTL the potential is replaced by
the sum of all inverse distances. Contrary to the KS-chain, zero masses 
do not cause any singularity in these AR-methods; TTL even provides equal 
weight to all the members of the subsystem, and thus these methods are well 
suited for integration of systems with large mass ratios.
Use of the generalized mid-point method \citep{MM-06} allows velocity-dependent
terms, e.g. the post-Newtonian expansion. 
In addition to the leapfrog and/or generalized mid-point method one must 
use the Bulirsch-Stoer (BS) extrapolation method for high accuracy. 
The basic algorithms provide the correct symmetry for the BS method to work 
efficiently.
In a recent implementation \citep{MM-08}, called AR-CHAIN,
the chain structure, introduced originally by \cite{MA-93},
is incorporated with a new time-transformation that combines the advantages of
the LogH and TTL and generalized mid-point methods. 
The chain structure significantly reduces the roundoff error and the other 
transformations provide regular data, necessary to achieve high precision, 
for the BS-extrapolator.
The AR-CHAIN code also includes post-Newtonian terms to order PN2.5

In this paper we describe the performance of a new, hybrid $N$-body 
code that incorporates AR-CHAIN.
The new code, called $\varphi$GRAPEch, is based on (the serial version of)
$\varphi$GRAPE, a general-purpose, direct-summation $N$-body code
which uses GRAPE special-purpose hardware to compute accelerations
\citep{harfst07}.
The new code divides particles into two  groups: particles
associated with the massive object (or objects) and that are included
in the chain, and particles outside the chain that are advanced
via the Hermite scheme of $\varphi$GRAPE.
Some of the latter particles are denoted as perturbers and are allowed
to affect the motion of stars in the chain, and vice versa.
After describing the hybrid code (\S\ref{sec:code}),
we present the results of performance tests based on a model
that mimics the star cluster around a supermassive black hole
(\S\ref{sec:perf}).
We show that the hybrid code can achieve higher overall accuracies
(as measured via energy conservation, say) than $\varphi$GRAPE alone,
and in less elapsed time, in spite of the additional overhead
associated with the chain.

In \S\ref{sec:appl} we apply $\varphi$GRAPEch to the integration of
a realistic, multi-component model of the Galactic center.
In the first application (\S\ref{sec:gc})
we accurately evaluate, for the first time, the
effects of perturbations from stars and stellar remnants on the
orbital elements of bright stars observed on short-period orbits
about Sagittarius A* \citep[S-stars:][]{Ghez-03,eisen-05}.
We then (\S\ref{sec:imbh}) present an integration of the orbit 
of an intermediate-mass
black hole as it spirals into the galactic center via the
combined influence of dynamical friction and gravitational-wave
energy loss.

\section{The Hybrid $N$-Body Code}
\label{sec:code}

In this section we describe how the AR-CHAIN algorithm of \citet{MM-08}
was integrated into the serial version of the direct-summation code
$\varphi$GRAPE \citep{harfst07}. 
The latter algorithm employs a Hermite
integration scheme \citep{MA92} with hierarchical, commensurate block
time steps and uses a GRAPE board to calculate forces.

We begin by briefly describing the integration scheme without the chain 
and its implementation using the GRAPE.

\subsection{Integration scheme}

In addition to position ${\bf x}_i$, velocity ${\bf v}_i$,
acceleration ${\bf a}_i$, and time derivative of acceleration ${\bf
\dot{a}}_i$, each particle $i$ has its own time $t_i$ and time step
$\Delta t_i$.

Integration consists of the following steps:
\renewcommand{\labelenumi}{(\arabic{enumi})}
\begin{enumerate}
\item
The initial time steps are calculated from
\begin{equation}
\Delta t_i = \eta_s\frac{|{\bf a}_{i}|}{|{\bf \dot{a}}_{i}|},
\label{eq:ts}
\end{equation}
where typically $\eta_s = 0.01$ gives sufficient accuracy.
\medskip
\item The system time $t$ is set to the minimum of all $t_i + \Delta t_i$,
and all particles $i$ that have  $t_i + \Delta t_i=t$ are selected as
active particles. 
\medskip
\item Positions and velocities at the new $t$ are predicted for all
particles using 
\begin{subequations}
\begin{eqnarray}
{\bf x}_{j,\mathrm{p}} &=& {\bf x}_{j,0} + (t-t_j) {\bf v}_{j,0} + 
\frac{(t-t_j)^2}{2}{\bf a}_{j,0} + \frac{(t-t_j)^3}{6}{\bf \dot{a}}_{j,0}, \\
{\bf v}_{j,\mathrm{p}} &=& {\bf v}_{j,0} +
(t-t_j){\bf a}_{j,0} + \frac{(t-t_j)^2}{2}{\bf \dot{a}}_{j,0}.
\end{eqnarray}
\end{subequations}
Here, the second subscript denotes a value given either at the
beginning (0) or the end (1) of the current time step. All quantities
used in the predictor can be calculated directly, i.e. no memory of a
previous time step is required.  
\medskip
\item Acceleration and its time derivative are updated for active
particles only according to
\begin{subequations}
\begin{eqnarray}
{\bf a}_{i,1} &=& \sum_{j\ne i} Gm_j\frac{{\bf
r}_{ij}}{(r_{ij}^2+\epsilon^2)^{(3/2)}}, \\
{\bf \dot{a}}_{i,1} &=& \sum_{j\ne i} Gm_j\left[\frac{{\bf
v}_{ij}}{(r_{ij}^2+\epsilon^2)^{(3/2)}} + \frac{3({\bf
v}_{ij}\cdot{\bf r}_{ij}){\bf r}_{ij}}{(r_{ij}^2+\epsilon^2)^{(5/2)}}\right],
\end{eqnarray}
\end{subequations}
where
\begin{subequations}
\begin{eqnarray}
{\bf r}_{ij} &=& {\bf x}_{j,p}-{\bf x}_{i,p}, \\
{\bf v}_{ij} &=& {\bf v}_{j,p}-{\bf v}_{i,p},
\end{eqnarray}
\end{subequations}
and $\epsilon$ is the softening parameter.
\medskip
\item Positions and velocities of active particles are corrected
using 
\begin{subequations}
\begin{eqnarray}
{\bf x}_{i,1} &=&  {\bf x}_{i,\mathrm{p}} 
               + \frac{\Delta t_i^4}{24}\myddot{a}_{i,0} 
               + \frac{\Delta t_i^5}{120}\mydddot{a}_{i,0}, \\ 
{\bf v}_{i,1} &=&  {\bf v}_{i,\mathrm{p}} 
               + \frac{\Delta t_i^3}{6}\myddot{a}_{i,0} 
               + \frac{\Delta t_i^4}{24}\mydddot{a}_{i,0}
\end{eqnarray}
\end{subequations}
where the second and third time derivatives of ${\bf a}$ are given by
\begin{subequations}
\begin{eqnarray}
\myddot{a}_{i,0} &=& \frac{-6\left({\bf a}_{i,0}-{\bf a}_{i,1}\right)
	                       - \Delta t_i 
                    \left(4{\bf \dot{a}}_{i,0}+2{\bf \dot{a}}_{i,1}\right)}
                           {\Delta t_i^2}, \\
\mydddot{a}_{i,0} &=& \frac{12\left({\bf a}_{i,0}-{\bf a}_{i,1}\right)
	                       +6 \Delta t_i 
                    \left({\bf \dot{a}}_{i,0}+{\bf \dot{a}}_{i,1}\right)}
                           {\Delta t_i^3}.
\end{eqnarray}
\end{subequations}
\medskip
\item The times $t_i$ are updated and the new time steps
$\Delta t_i$ are determined.
Time steps are calculated using the standard formula \citep{A85}:
\begin{equation}
\Delta t_{i,1} = \sqrt{\eta\displaystyle\frac{|{\bf a}_{i,1}||
\myddot{a}_{i,1}| + |{\bf \dot{a}}_{i,1}|^2}
{|{\bf \dot{a}}_{i,1}||\mydddot{a}_{i,1}| + |\myddot{a}_{i,1}|^2}}.
\label{eq_timestep}
\end{equation}
The parameter $\eta$ controls the accuracy of the integration and is
typically set to $0.01$ (although the use of smaller value of
$\eta$ is described below).
The value of $\myddot{a}_{i,1}$ is calculated from 
\begin{equation}
\myddot{a}_{i,1} = \myddot{a}_{i,0} + \Delta t_{i,0}\mydddot{a}_{i,0}
\end{equation}
and $\mydddot{a}_{i,1}$ is set to $\mydddot{a}_{i,0}$.
\medskip
\item Repeat from step (2).
\end{enumerate}

A hierarchical commensurate
block time step scheme is necessary when the Hermite integrator
is used with the GRAPE (and is also efficient for parallelization and
vectorization; see
below and \citet{mcmillan-86}). 
Particles are grouped by replacing their time steps $\Delta
t_i$ with a block time step $\Delta t_{i,\mathrm{b}}=(1/2)^n$, where
$n$ is chosen according to
\begin{equation}
\left(\frac{1}{2}\right)^n \le \Delta t_i < \left(\frac{1}{2}\right)^{n-1}.
\end{equation} 
The commensurability is enforced by requiring that $t/\Delta t_i$ 
be an integer.
For numerical reason we also set a minimum time step $\Delta t_\mathrm{min}$,
where typically
\begin{equation}
\Delta t_\mathrm{min} = 2^{-m},
\label{eq_tmin}
\end{equation}
with $m \ge 28$. The time steps of particles with $\Delta t_i < \Delta
t_\mathrm{min}$ are set to this value. The minimum time step should be
consistent with the maximum acceleration defined by the softening
parameter; monitoring of the total energy can generally indicate
whether this condition is being violated.

\subsection{GRAPE implementation}
\label{sec_G6impl}

The GRAPE-6 and GRAPE-6A hardware has been designed to work with a
Hermite integration scheme and is therefore easily integrated into the
algorithm described in the previous section \citep[see][]{MFK03}. 
In detail, integration of particle positions using the GRAPE-6A
consists of the following steps:

\begin{enumerate}
\item \textbf{Initialize} the GRAPE and send particle data (positions, velocities,
etc.) to GRAPE memory. 
\\
\item \textbf{Compute} the next system time $t$ and select active particles on
the host (same as step 2 in previous section).
\\
\item \textbf{Predict} positions and velocities of active particles only and
send the predicted values together with the new system time $t$ to
GRAPE's force calculation pipeline.
\\
\item \textbf{Predict} positions and velocities for all other
particles on the GRAPE, and calculate forces and their time 
derivatives for active particles.
\\
\item \textbf{Retrieve} forces and their time derivatives from the
GRAPE and \textbf{correct} positions and velocities of active particles
on the host.
\\
\item \textbf{Compute} the new time steps and update the particle data
on the host of all active particles in the GRAPE memory. 
\\
\item \textbf{Repeat} from step (2).
\end{enumerate}

\subsection{Integrating AR-CHAIN into $\varphi$GRAPE}

Here we describe the hybrid $N$-body code.
We first describe the step-by-step algorithm as in the previous section,
then provide a more detailed explanation of the most important steps.

\begin{enumerate}
\item \textbf{Initialize} the GRAPE and send particle data (positions,
velocities, etc.) to GRAPE memory.  
\\
\item \textbf{Compute} the next system time $t$ and select active
particles on the host (same as step 2 in previous section).  \\

\item If the chain is not active, \textbf{check} the active particles 
to see whether the chain should be used. 
If the chain is active, \textbf{check} if particles
enter or leave the chain.  
\\
\item If the chain is not needed, standard integration can continue
from step (14). 
Otherwise, \textbf{(re-)initialize} the chain if a new chain
has been created or an existing chain has changed.  
\\
\item \textbf{Find} all particles within a sphere of radius
$r_\mathrm{pert}$ centered on the center-of-mass (COM) chain particle
and define them as perturbers.
\\
\item  \textbf{Compute} forces on the COM particle (if the chain is
new or has changed).
\\
\item \textbf{Evolve} the chain one time step.
\\
\item \textbf{Predict} the position and velocity of the COM chain
particle, \textbf{compute} forces and \textbf{correct} its 
position and velocity.
\\
\item \textbf{Resolve} the chain and \textbf{update} positions and
velocities of both the chain COM particle and the
individual chain particles.  
\\
\item \textbf{Integrate} all active particles; 
particles within $r_\mathrm{res}$ feel the forces from the resolved chain 
(i.e.\ individual chain particles) and particles outside 
$r_\mathrm{res}$ feel only the forces from the chain COM particle. 
\\   
\item \textbf{Compute} new time steps for the active particles.
\\
\item \textbf{Update} the particle data on the host CPU
and the GRAPE for all active particles.
\\
\item \textbf{Repeat} from step (2).
\\ 
\item \textbf{Predict} positions and velocities of active particles only and
send the predicted values together with the new system time $t$ to
the GRAPE force calculation pipeline.
\\
\item \textbf{Predict} positions and velocities for all other
particles on the GRAPE, and calculate forces and their time 
derivatives for active particles.
\\
\item \textbf{Retrieve} forces and their time derivatives from the
GRAPE and \textbf{correct} positions and velocities of active particles
on the host.
\\
\item \textbf{Compute} the new time steps and update the particle data
on the host of all active particles in the GRAPE memory. 
\\
\item \textbf{Repeat} from step (2).
\end{enumerate}

In the remainder of this section, 
we describe in detail the implementation of the hybrid code. 
For purposes of discussion, we assume a test case where a very massive
particle is located near the center of the system, e.g.\ a 
supermassive black hole in the galactic center. 
But the same algorithm can be easily applied to other
configurations as well.

The hybrid code is initialized in the manner  described above. 
After the time steps for all particles have been determined, 
a check is made whether to use the chain. 
Two parameters control the assignment of particles to the chain:
$t_\mathrm{crit}$ and $r_\mathrm{crit}$. 
Initially, any particle that satisfies both criteria:
\begin{subequations}
\begin{eqnarray}
  \Delta t_i &\le& t_\mathrm{crit} \\
\label{eq:tcrit}
  \Delta r_{i,\mathrm{BH}} &\le& r_\mathrm{crit} 
\label{eq:rcrit}
\end{eqnarray}
\end{subequations}
\noindent
is selected as a chain particle, where $r_{i,\mathrm{BH}}$ is the
distance of particle $i$ to the massive particle. 
Once the chain is activated and a chain radius (as defined
below) is determined,
all particles within $r_\mathrm{ch}$ are assigned to the chain,
i.e. $t_{\rm crit}$ is ignored.
If two or more particles are selected as chain particles the chain is
activated; the massive particle is always designated a chain particle. 

If the chain is not needed the code continues with the standard
$N$-body integration. Otherwise a new chain is created (step 4). 
The chain radius $r_\mathrm{ch}$ is set to the largest
$r_{i,\mathrm{BH}}$ found in the previous step and all particles within
$r_\mathrm{ch}$ are added to chain independent of their time step. Any
chain particle with $t_i \ne t_\mathrm{sys}$, where $t_\mathrm{sys}$
is the current system time, is integrated to this time. Then a pseudo
chain particle is created with the COM coordinates of the
chain particles. The chain particles are also removed from the list of
active particles and their masses are set to zero on the GRAPE.

All particles within $r_\mathrm{pert}$ are selected as perturber
particles (5) and the perturber radius is calculated from
\begin{equation}
r_\mathrm{pert} = \left(\frac{m_*}{F M_\mathrm{ch}}\right)^\frac{1}{3}
\times 1.5 r_\mathrm{crit},
\end{equation}
where $m_*$ is the mass of a particle and $M_\mathrm{ch}$ the total
mass in the chain. The parameter $F$ controls the accuracy of the
calculation of forces on the chain. Typically, $F = 10^{-6}$ is
used. In principle $r_\mathrm{pert}$ should be proportional to
$r_\mathrm{ch}$. However, $r_\mathrm{ch}$ can sometimes change
significantly during a chain step and it turned out that using $1.5
r_\mathrm{crit}$ results in a much smaller integration error (the
additional factor of $1.5$ is due to the condition for particles to leave
the chain). It also allows the perturber list to be re-used for several
steps making the algorithm more efficient. Perturber particles act on
the resolved chain as described below.

The force on the COM particle is calculated (6) in
two steps. First, the force of the non-perturbers (i.e.\ particles
outside of $r_\mathrm{pert}$) on the COM particle is computed. This
is done by loading the COM particle to the GRAPE and setting the
masses of all perturber particles to zero at the same time. 
Then, a standard GRAPE call is used to compute the force, and
the masses of the perturbers are set back to normal. 
In the second step, the forces of
the perturbers on each chain particle are computed on the host. 
(In our applications, the number of perturbers is usually less than $100$.
Some problems may require a much larger number of perturbers, in which case
it would be more efficient to use the GRAPE to do this force
calculation as well.) 
The force on the individual chain particles is summed up according to
\begin{equation}
{\bf a}_\mathrm{COM,pert} = \frac{1}{M_\mathrm{ch}} \sum_{i_\mathrm{ch}}^{N_\mathrm{ch}} m_{i_\mathrm{ch}}{\bf a}_{i_\mathrm{ch}}
\end{equation}
to give the total force on the COM particle (and likewise for the
force derivatives).

In the next step (7), the chain particles are advanced for one system
time step (which is the shortest time step needed by a non-chain
particle and ideally longer than the shortest regular time step of the
chain particles). 
This is achieved by sending the masses, positions and
velocities of the chain and perturber particles plus the forces and
force derivatives of the perturbers to AR-CHAIN. The COM of
the chain can change during this step because the chain is perturbed
but any change is subtracted at this point.

Now a predictor-corrector step is made for the COM particle 
(step 8). 
Using the force calculated in step (6), position and velocity are
predicted with equation~(2), and these are then used to 
recompute the force on the COM particle in the same way 
as described above.
The corrected position and velocity can then be calculated by equation~(5).

The new position of the COM particle, together with the new positions
and velocities of the chain particles calculated in the chain (7) can
be used to resolve the chain, thus the true positions and velocities
at the end of the current time step can now be written to the memory
and GRAPE (including the COM particle).  With the resolved chain it is
now possible to compute the forces for the active particles. This
again is done in two steps for particles outside and inside of
$r_\mathrm{res} = F^{-1/3}r_\mathrm{ch}$: First all active particles
outside of $r_\mathrm{res}$ are integrated with the chain particles
replaced by the COM particle. Then the active particles inside
$r_\mathrm{res}$ are integrated seeing the resolved chain and not the
COM particle. This is done by removing the COM particle from the GRAPE
and by setting the masses of the chain particles back to normal.

Finally, a new time step can be computed for all active particles
(note that the chain particles are not counted as active but
everything in the chain is integrated in every step nonetheless). 
Positions and velocities are also updated on both the
host and the GRAPE. Then, the integration can continue with the next
step (going back to (2)).

If the chain is already active at the beginning of the time step,
non-chain particles are checked to see whether they have entered the chain. 
A particle enters the chain if its distance to the chain COM is smaller
than the chain radius $r_\mathrm{crit}$. Any particle that enters the
chain is, if needed, syncronized to $t_\mathrm{sys}$ first. Particles
leave the chain if their distance to the COM becomes larger than
$1.5r_\mathrm{crit}$. The factor of $1.5$ ensures that particles do
not enter and leave the chain too often (to avoid the overhead
associated with extra internal initialization
each time chain membership is changed).
A particle leaving the
chain has its force, derivative and time step initialized and this
information is sent to the GRAPE when the mass of this particle is set
back to normal. 
The initialization of the time step is particularly
important: errors are easily introduced into the integration if the
time step is chosen too large. 
We use Eq.~(\ref{eq:ts}) with a rather small value for $\eta_s$, 
e.g.\ $\eta_s = 10^{-2}\eta$.

The search of perturber particle in step (5) is computationally rather
expensive and is therefore not done in every step. A parameter $\Delta
t_\mathrm{pert}$ is used to determine how often the list of perturbers
is renewed (note that only the list of perturbers is unchanged for
some time, current position etc.\ are still used within the chain). In
addition, the perturber list is renewed whenever chain membership is
changed. Also, the force on the COM particle only needs an update if
the chain membership has changed.

Finally, an extra GRAPE call is needed to compute the potential for
chain particles if at the end of a step if output of energies, etc. is
required.

\section{Results of Performance Tests}
\label{sec:perf}

\begin{figure}
  \begin{center}
  \includegraphics[width=8cm]{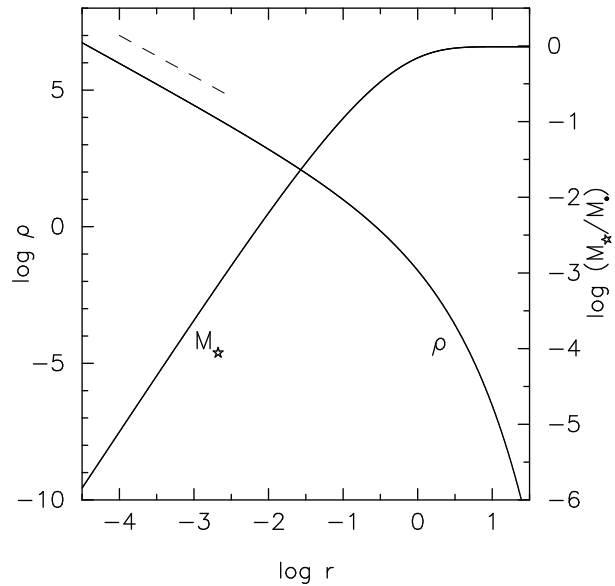}
  \caption{Density and cumulative mass profiles of the model used in the
    performance tests.  The density obeys $\rho\propto r^{-1.5}$ near
    the center (dashed line) and is truncated via Einasto's law beyond
    $r = 0.1$.  The total mass in stars equals that of the
    ``black hole'' particle.  This model can be roughly scaled to the
    Galactic center star cluster by setting the length unit to $1\pc$
    and the total stellar mass to $3\times 10^6\msun$.}
  \label{fig:model}
  \end{center}
\end{figure}

We tested the performance of the hybrid code using various realizations
of a model designed to mimic the density profile of the star cluster 
around the Milky Way supermassive black hole (Fig.~\ref{fig:model}).
The model has a mass density profile
\begin{equation}
\rho(r) = \rho_0 \left(\frac{r}{R_e}\right)^{-3/2} 
\exp\left[{-b\left(r/R_e\right)^{1/n}}\right].
\end{equation}
This is a $\rho\sim r^{-3/2}$ power-law near the center, similar to
what is observed in the inner $\sim 0.1\pc$ of the Milky Way
\citep{Schoedel-07}.  
An Einasto-like truncation was applied to give the
model a finite total mass; the Einasto index was $n=2$ and we adjusted
$b$ such that the truncation begins at a radius of $\sim 0.1 R_e$.  The
total mass in stars was fixed to be equal to that of the BH particle,
i.e. all ``star'' particles have mass $m=M_\bullet/N$.  Henceforth units
are adopted such that $G=M_\bullet=R_e=1$; the model can be scaled
approximately to the Galactic center by setting the length unit to $1\pc$
and the mass unit to $3\times 10^6 \msun$, making the unit of time
$\sim 10^4$ yr.  

Initial positions and velocities for the $N$ "star"
particles were generated as follows. (1) Eddington's formula was
used to compute the unique, isotropic phase-space distribution function 
$f_i(E)$, $i=1,...,4$ that reproduced the adopted $\rho_i(r)$ of each 
species in the combined gravitational potential of the BH and the stars. 
(2) Distances $r$ from the BH were generated by sampling randomly from 
the integrated mass profiles, and ($x,y,z$) coordinates were assigned by 
selecting random positions on each sphere of radius $r$.
(3) The (isotropic) velocity distribution at this radius was computed
from $\Phi(r)$ and $f(E)=f[v^2/2+\Phi]$.
(4) The magnitude of the velocity was selected randomly from this function,
and the Cartesian velocity components were generated in a manner analogous 
to the position coordinates.
Unless otherwise specified, performance tests were
carried out on single nodes of {\tt gravitySimulator}, a 32-node
cluster with a GRAPE6-A card on each node \citep{harfst07}.

\begin{figure}
  \begin{center}
    \includegraphics[width=8cm]{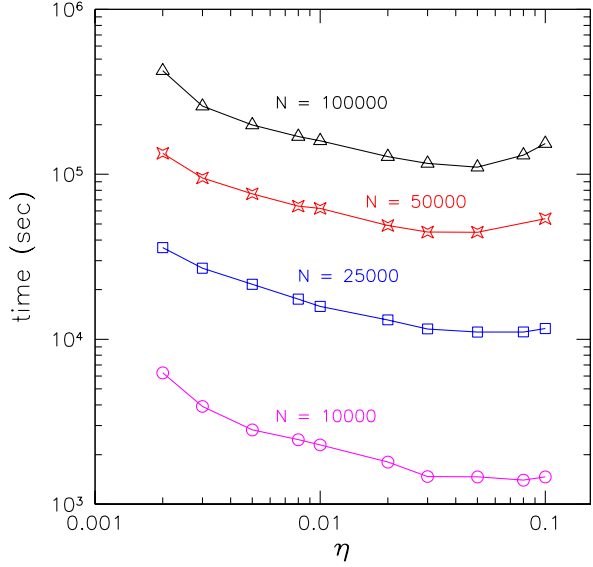}
    \caption{Elapsed time as a function of $N$ and $\eta$ in
      integrations until time $t=1$, and with $r_{\rm crit}=
      4\times10^{-4}$.  }
  \label{fig:Ncfr}
  \end{center}
\end{figure}
Figure~\ref{fig:Ncfr} shows the integration time as a function of
$\eta$ for various $N$ and for fixed $r_{\rm crit}=4\times10^{-4}$.
For this value of $r_{\rm crit}$, the typical number of particles in
the chain, at any given time, is roughly linear in the total particle
number. The number of particles in the chain is also small enough
so that the total execution time is dominated by integration of
particles outside the chain. This can be seen in Fig.~\ref{fig:Ncfr}
which shows that the scaling with the number of particles is
approximately $N^2$ (the scaling in the chain-dominated case is
discussed below).

\begin{figure}
 \begin{center}
    \includegraphics[width=8cm]{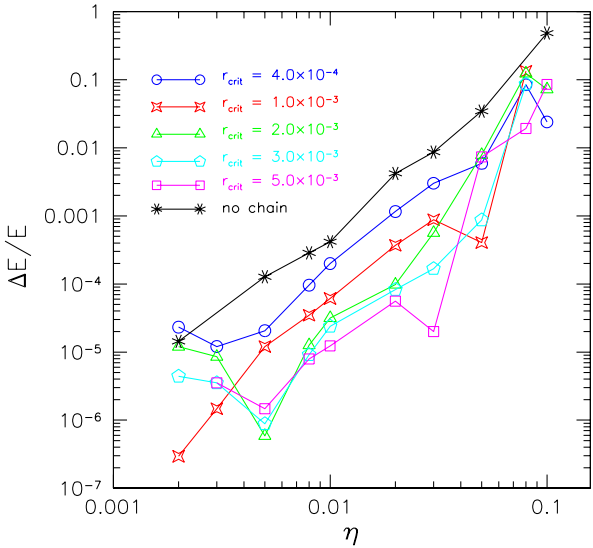}
  \caption{Energy conservation in integrations until $t=1$ ($\sim
    10^4$ yr) of the model illustrated in Fig.~\ref{fig:model} with
    $10^4$ particles.  Black line (asterisks) are for $\varphi$GRAPE
    without the regularized chain.  }
  \label{fig:ene}
\end{center}
\end{figure}

\begin{figure}
 \begin{center}
  \includegraphics[width=8cm]{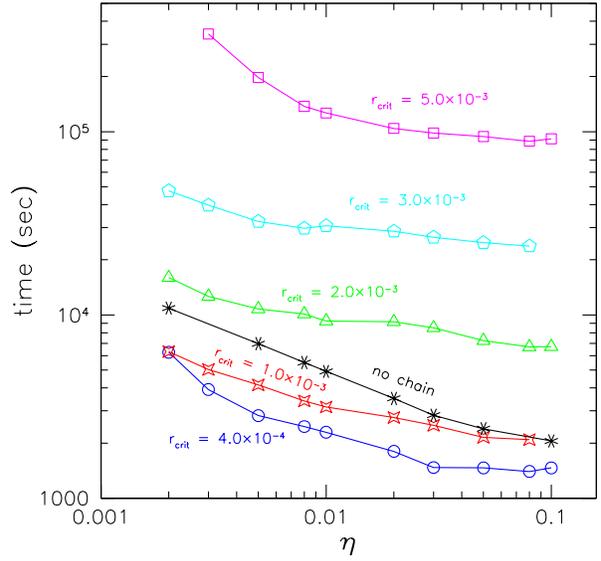}
  \caption{Elapsed time for the integrations of Fig.~\ref{fig:ene}.}
  \label{fig:time}
\end{center}
\end{figure}
Figures~\ref{fig:ene} and~\ref{fig:time} show the energy conservation
and elapsed time for integrations until $t=1$ for the case $N=10^4$
and for various values of $\eta$, the accuracy parameter in the
Hermite integrator (eq.~\ref{eq:ts}), and $r_{\rm crit}$, the maximum
distance from the black hole at which a particle enters the chain
(eq.~\ref{eq:rcrit}); $t_{\rm crit}$ (eq.~\ref{eq:tcrit}) was fixed at
$5\times10^{-5}$, the softening parameter was $\epsilon =
  10^{-5}\pc$ and post-Newtonian terms were not included.  (Including
the PN terms was found to affect the speed of the code only very
slightly; they were omitted in order to simplify the discussion of
energy conservation.)  Also shown is the performance of $\varphi$GRAPE
without the regularized chain.  The figures show the expected scaling
of the Hermite scheme with the accuracy parameter: time steps increase
linearly with $\eta$ making the integration faster and less accurate.
For a fourth-order scheme, the relative energy error scales as $\sim
dt^5\sim \eta^{5/2}$, though in the case of the hybrid code the
relation is modified by the presence of the chain.  Energy
conservation generally improves for larger values of $r_{crit}$, as
more and more particles are removed from the $N$-body integration and
are treated more accurately in the chain. The integration time
increases rapidly with $r_{\rm crit}$ reflecting the $\sim n_{\rm
  ch}^3$ dependence of the chain.  Nevertheless it is clear that for
{\it all} values of $\eta$, there exist values of $r_{\rm crit}$ such
that the hybrid code is both more accurate, {\it and faster}, than
$\varphi$GRAPE alone.  This is presumably because the additional
computational overhead associated with the regularization scheme is
more than compensated for by the longer mean time steps of particles
outside the chain.

\begin{figure}
  \begin{center}
    \includegraphics[width=8cm]{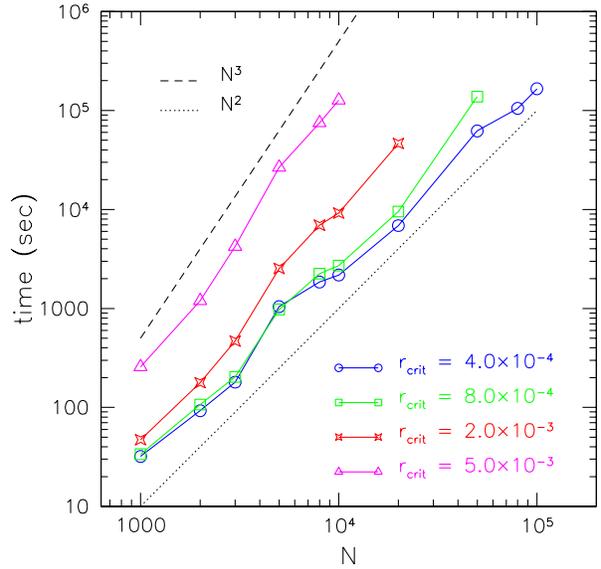}
    \caption{$N$-dependence of the elapsed time for integrations until
      $t=1$ of the model in Figure~\ref{fig:model}, with $\eta=0.01$.  As
      the mean number of particles in the chain increases, the
      $N$-scaling changes from $\sim N^2$ (dotted line) to $\sim N^3$
      (dashed line).  }
  \label{fig:chaintime}
  \end{center}
\end{figure}
The $N$-scaling of the integration time is more apparent in
Fig.~\ref{fig:chaintime}.  When $r_{\rm crit}$ is small, most of the
computation is spent on the non-chain particles and the performance
scales as $\sim N^2$.  As $r_{\rm crit}$ increases, so does the
typical number $n_{\rm ch}$ of particles in the chain, and the
$N$-scaling of the total integration time changes to $\sim n_{\rm
  ch}^3\sim N^3$.  Along the line $r_{\rm crit}=2\times 10^{-3}$ in
Figure~\ref{fig:chaintime}, which roughly marks the transition from an
$\sim N^2$ to an $\sim N^3$ scaling, the mean fraction $n_{\rm ch}/N$
of particles in the chain varies from $\sim 3\times 10^{-4}$ to $\sim
7\times 10^{-4}$.  If we postulate a performance model in which the
total integration time is simply the sum of the time spent on
particles in the chain, $t_{\rm ch} = An_{\rm ch}^3$, plus the time
spent on particles outside the chain, $t_N=BN^2$, then the fraction
$(n_{\rm ch}/N)_{\rm crit}$ at which the chain begins to dominate the
total time should scale as $N^{-1/3}$.  We did not attempt to verify
this prediction in detail, but if we adopt $(n_{\rm ch}/N)_{\rm
  crit}\approx 5\times 10^{-4}$ at $N=10^4$, the model predicts
$(n_{\rm ch}/N)_{\rm crit}\approx 0.01 N^{-1/3}$.  This relation can
be taken as defining the effective upper limit to the number of
particles to include in the chain; for $N=10^6$, $n_{\rm ch,
  crit}\approx 100$.

\begin{figure}
  \begin{center}
    \includegraphics[width=8cm]{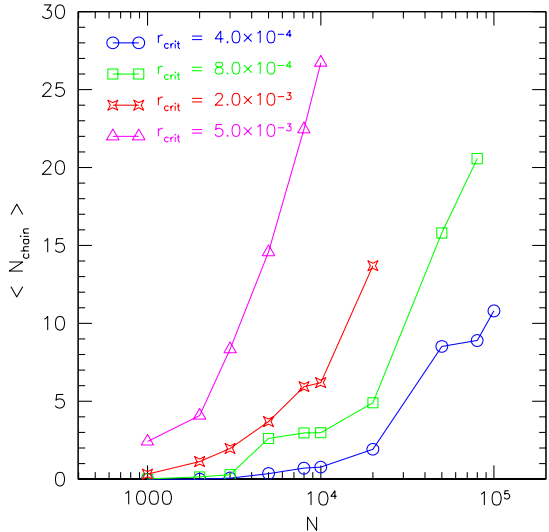}
    \caption{Average number of chain particles for integrations until
      $t=1$ of the model in Figure~\ref{fig:model}, with $\eta=0.01$.}
  \label{fig:nchain}
  \end{center}
\end{figure}
The average fraction of chain particles in the reported integrations
spans the range $(3-7)\times 10^{-4}$. This reflects a $N$-dependence
of the average number of chain particles, as can be seen in
Fig.~\ref{fig:nchain}. In these test runs, we set a hard maximum
number of chain particles of $n_{\rm ch, crit} = 50$.

The results presented in this section constitute a particularly severe
test of the code, since all of the particles are moving, at all times,
in the essentially Keplerian potential of the central mass.  In many
other applications, the sphere of influence of the massive particle(s)
would only include a fraction of the other particles in the
simulation, and a choice of $r_{crit}$ could be made that included all
the particles in this region without forcing $n_{ch}$ to unreasonable
values.

\section{Applications}
\label{sec:appl}
\subsection{Evolution of Orbits near the Galactic Center Black Hole}
\label{sec:gc}

AR-CHAIN conserves the Keplerian elements of unperturbed two-body
orbits with extremely high precision, even for arbitrarily large mass
ratios, and this fact makes the hybrid code uniquely suited to
investigating the detailed effects of perturbations on the orbits of
individual stars around the galactic center supermassive black hole
(SMBH).

We illustrate this using the collisionally relaxed, multi-mass model
of \cite{HA-06a} (hereafter the ``HA06 model'').  This model has four
components in addition to the SMBH: main sequence (MS) stars,
$m=1\msun$; white dwarfs (WD), $m=0.6 \msun$; neutron stars (NS),
$m=1.4\msun$; and stellar-mass black holes (BH), $m=10\msun$.  The
HA06 model was derived as a steady-state solution of the isotropic,
orbit-averaged Fokker-Planck equation assuming a SMBH mass of $3\times
10^6 \msun$, and including an approximate term representing loss of
stars into the tidal disruption sphere.  The contribution of the stars
to the gravitational potential was ignored, making the solution valid
only within the SMBH's influence radius, $r\lap 1\pc$.  The three
lighter species (which dominate the density at most radii) have $\rho
\sim r^{-\gamma}$, $1.4\lap\gamma\lap 1.5$, $r\gap 3\mpc$ while the
heavier BHs have a steeper profile, $\gamma\approx 2$.  Detailed
density profiles for the four species were kindly provided by
T. Alexander.  We modified the HA06 model by imposing a steep
truncation like that of Figure~\ref{fig:model} to the density of each
species beyond $r=0.1\pc$, then computed the self-consistent isotropic
phase-space density $f_i(E), i=1,...,4$ corresponding to each species
from the truncated $\rho_i(r)$ profiles using Eddington's formula.
Finally, Monte-Carlo positions and velocities were generated from the
$\rho_i$ and $f_i$.  The total number of objects of all types was
found to be $\sim 75,000$, with $\sim 10^3$ objects (mostly MS stars)
within $0.01\pc$; the latter number is consistent with the value given
in Table 1 of \cite{HA-06a}.  Hereafter, we refer collectively to
stars and stellar remnants in the $N$-body simulation simply as
``stars.''

The non-zero masses of the stars cause the $N$-body gravitational
potential to deviate slightly from the fixed Keplerian potential of
the SMBH, and the resultant perturbations cause the orbital elements
of any single (``test'') star to evolve.

As test stars, we included five particles with orbital elements
corresponding to the five, shortest-period S-stars observed near the
galactic center: S0-1, S0-2, S0-16, S0-19, and S0-20. The S-stars
\citep{Ghez-03,eisen-05} are a cluster of main sequence, O-B stars 
that orbit around Sagittarius A* with periods as short as 15 yr.
Their short periods and their proximity to the supermassive black hole 
make them very interesting objects, e.g.\ for constraining the mass 
of the black hole
or studying the effects of dynamical and/or relativistic 
perturbations on their orbits.
In our model, the masses of these five S-stars were set to
$15\msun$ and initial positions and velocities were determined at year
2000 AD using the Keplerian orbital elements given in Table 3 of
\cite{ghez-05}.  (Initial velocities of the S-stars were adjusted, at
fixed $a$ and $e$, to account for the slightly different values of
$\mbh$ assumed by Hopman \& Alexander (2006) and Ghez et al. (2005).)
These stars are being constantly monitored and deviations of their
orbits from closed Keplerian ellipses might be used to test various
hypotheses about the distribution of matter near the SMBH, or as a
test of general relativity
\citep[e.g.][]{fragile-00,rubilar-01,weinberg-05,zucker-06,will-08}.

\begin{figure}
  \begin{center}
    \includegraphics[width=8cm]{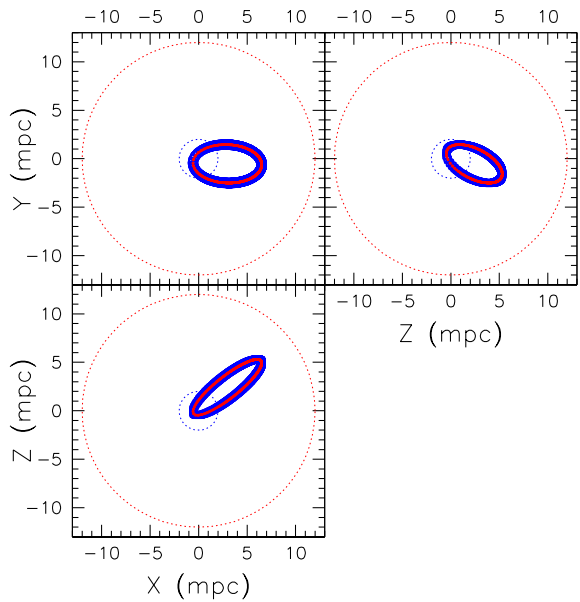}
  \end{center}
  \caption{Orbit of the galactic center star S0-2 over a time of 100
    yr in the case of $r_{\rm crit} = 2\mpc$ (blue) and $r_{\rm
      crit} = 12\mpc$ (red). The total number of stars in these
    integrations was $10^3$ and post-Newtonian terms were not included.}
  \label{fig:s2orb}
\end{figure}

\begin{figure}
  \begin{center}
    \includegraphics[width=8cm]{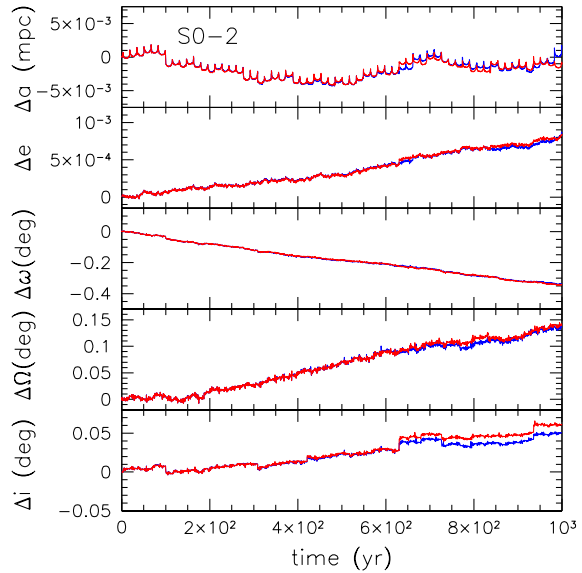}
  \end{center}
  \caption{Evolution of the orbital elements of star S0-2 in
    integrations with $N=10^3$ stars and $r_{\rm crit} = 2\mpc$
    (blue) and $r_{\rm crit} = 12\mpc$ (red).}
  \label{fig:s2kpl}
\end{figure}

The total number of stars in the HA06 model contained within the
S-star orbits is $\gap 10^3$, too large for all of them to be included 
in the chain at one time; this necessitates a choice of $r_{\rm crit}$ 
such that the S-stars will pass in and out of the chain in each orbit.  
We first verified that passage through $r_{\rm crit}$ did not in itself
introduce significant changes in the orbital elements.
Figures~\ref{fig:s2orb} and~\ref{fig:s2kpl} show the results of one such
set of tests, which followed the orbit of S0-2 for two different values of
$r_{\rm crit}$: $r_{\rm crit} = 2\mpc$ and $r_{\rm crit} = 12\mpc$,
compared with the semi-major axis length of $4.5\mpc$.  The total number of
stars in this test was set to $10^3$ in order that all particles
within the largest $r_{\rm crit}$ could be included in the chain
without exceeding a chain membership of $100$.  
Post-Newtonian terms were not included.
Figure~\ref{fig:s2kpl} shows the evolution of the five classical
elements of the Keplerian orbit of S0-2 over a time span of $10^3$ years.
There are only slight differences between the two runs, 
verifying that entrance into, or departure from, the chain
does not significantly influence the orbit.

\begin{figure}
  \begin{center}
    \includegraphics[width=8cm]{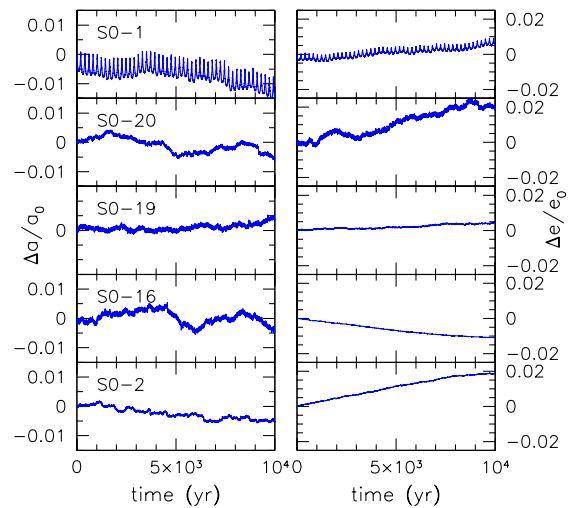}
  \end{center}
  \caption{Evolution of the semi-major axis and eccentricity variation
    for the five S-stars in an integration with $N=7.5\times 10^4$ stars 
    and $r_{\rm crit} = 0.8\mpc$.}
  \label{fig:ssae}
\end{figure}
The evolution of the semi-major axis and eccentricity of the all the
S-stars is shown in Fig.~\ref{fig:ssae} for an integration with
$N=7.5\times 10^4$ stars and $r_{\rm crit} = 0.8\mpc$.  While the
eccentricity evolves more or less linearly with time for all the
S-stars, the semi-major axis shows essentially a random evolution.

The timescale for apoastron precession in the S-stars 
(represented by $\Delta\omega$ in Fig.~\ref{fig:s2kpl})
is $\gap 10^5$ yr \citep[e.g.][]{weinberg-05}.
On shorter timescales, the angular momentum of stars like S0-2 
should evolve approximately 
linearly with time due to the (essentially fixed) torques resulting from  
finite-$N$ departures of the overall potential from spherical
symmetry \citep{RT-96}.

\begin{figure*}
  \begin{center}
    \includegraphics[width=15cm]{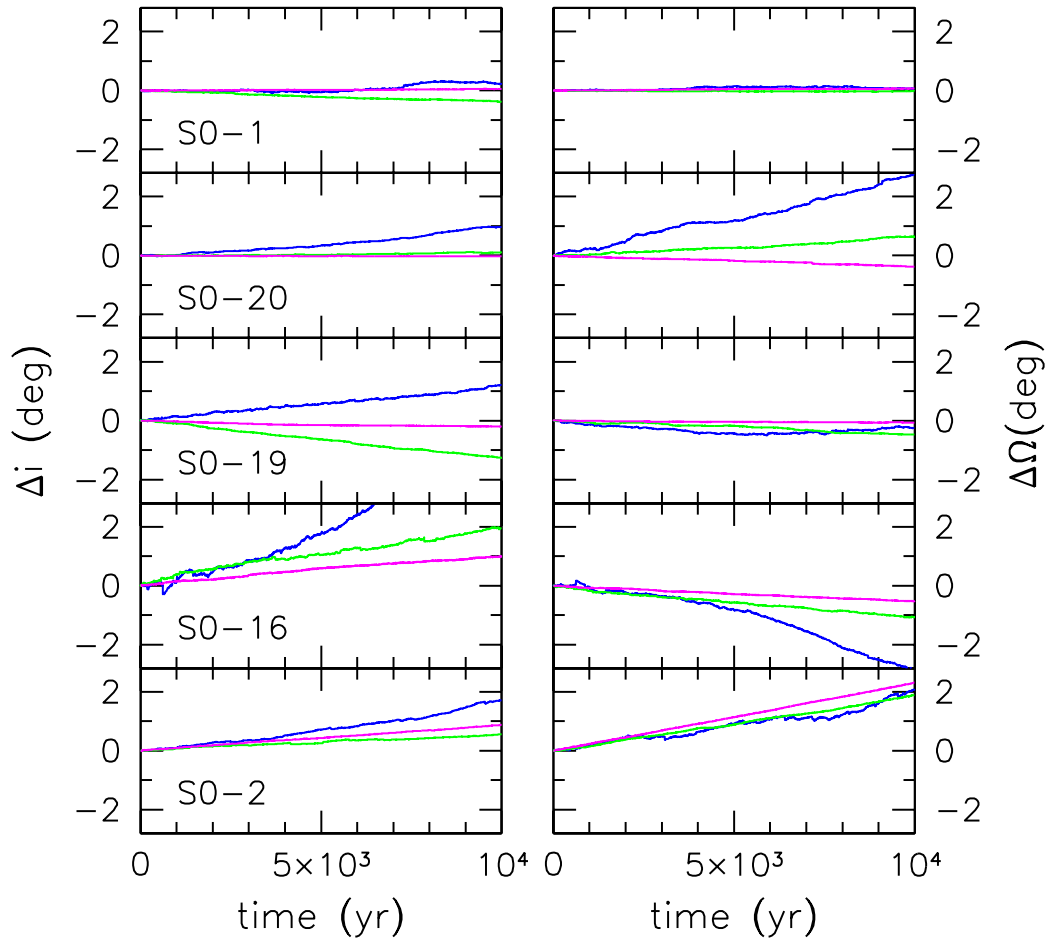}
  \end{center}
  \caption{Evolution of the orbital plane of the S-stars over a time
    span of $10^4$ years in integrations with different particle
    numbers: $N=10^3$ (magenta), $N=10^4$ (green), $N=7.5\times10^4$
    (blue). The left panels show the inclination angle $i$ while the
    right panels show position angle of the nodal point $\Omega$.}
  \label{fig:angles}
\end{figure*}

This evolution is illustrated for all the S-stars in
Figure~\ref{fig:angles}. 
The blue lines in that figure are from an integration that 
used the complete set of $N=7.5\times 10^4$ stars in the 
Monte-Carlo realization of the HA06 model,
and $r_{\rm crit} = 0.8\mpc$;
also shown are integrations that used randomly-chosen
subsets of $10^4$ stars ($r_{\rm crit}=2\mpc$) 
and $10^3$ stars ($r_{\rm crit}=10\mpc$) from this model.
Plotted in Figure~\ref{fig:angles} are the two Keplerian elements
($i,\Omega$) = (inclination, right ascension of ascending node) 
that measure the orientation of the orbital planes.
These angles would remain precisely constant in any spherical
potential and their evolution is due entirely to finite-$N$ departures 
of the potential from spherical symmetry.
The two angles are related to the Cartesian components of the
angular momentum via
\begin{subequations}
\begin{eqnarray}
L_x &=& L\sin i \sin\Omega, \\
L_y &=& -L\sin i \cos\Omega, \\
L_z &=& L\cos i.
\end{eqnarray}
\end{subequations}

Simple arguments \citep{RT-96} suggest that orbital inclinations
should evolve in this regime approximately as
\begin{subequations}
\begin{eqnarray}
\Delta\left(i,\Omega\right) &\approx& A \frac{m}{\mh} N^{1/2}
\frac{t}{P} \\
\label{eq:RR1}
&\approx& A \frac{m}{2\pi} \left(\frac{GN}{\mh a^3}\right)^{1/2} t
\label{eq:RR2}
\end{eqnarray}
\label{eq:RR}
\end{subequations}
where $m$ is a typical perturber mass, $N$ is the number of stars
within a sphere of radius $a$, the semi-major axis of the test star,
and $P(a)$ is the (Keplerian) orbital period.  The coefficient $A$ is
of order unity and is believed to depend weakly on orbital
eccentricity \citep{RT-96, gurkan07}.  We evaluated
equation~(\ref{eq:RR2}) numerically for the $N=75K$ model and found
that the dominant contribution to the torques is predicted to come
from the BH particles; the predicted change in orientation of a test
star over $10^4$ yr, for $A=1$, is $\sim 0.5^\circ$ for $a=10\mpc$
increasing to $\sim 1^\circ$ for $a=2\mpc$.  This is quite consistent
with Figure~\ref{fig:angles} if $2\lap A \lap 3$. We also do not see
an obvious contradiction with the results of \citet{gurkan07} with
regard to the eccentricity dependence of $A$. However, a more detailed
study is needed to give a definite answer to this question.  The
dependence of the evolution on $N$ in Figure~\ref{fig:angles} is also
consistent with the $N^{1/2}$ prediction of equation~(\ref{eq:RR}).

\subsection{Inspiral of an IMBH into the Galactic Center}
\label{sec:imbh}

As a second test problem, we used $\varphi$GRAPEch to follow the
inspiral of an intermediate-mass black hole (IMBH) into the Galactic
SMBH.  The multi-mass stellar cusp model described in the previous
sub-section was again used, with $N=75K$.  The second black hole was
given a mass of $10^{-3}$ times that of the SMBH, or $3\times
10^3\msun$; its initial orbit around the SMBH had semi-major axis
$0.1\mpc$ and its eccentricity was $0.9$.  This initial separation is
of the same order as the so-called hard-binary separation $a_h$ at
which inspiral due to dynamical friction alone would be expected to
stall \citep[e.g.][eq. 4.1]{GM-08}.  The large eccentricity was chosen
primarily to accelerate the inspiral.  
The integration used $r_{\rm crit}=0.8\mpc$, $\eta=0.01$ and
$\epsilon=10^{-5}\mpc$ and required a time of $\sim 60$ hr on one node of the
GRAPE cluster.

\begin{figure}
  \begin{center}
    \includegraphics[width=8cm]{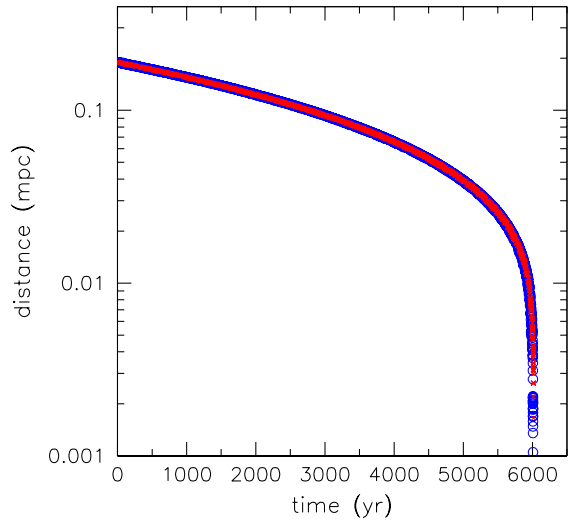}
  \end{center}
  \caption{Evolution of the distance between the IMBH and the SMBH
    over the inspiral time. The blue points refer to the simulation
    containing $N=75K$ stars while the red points refer to
    a simulation of the black hole binary in isolation.}
  \label{fig:dist}
\end{figure}

\begin{figure}
  \centering
  \begin{minipage}[b]{6 cm}
    \includegraphics[width=5cm]{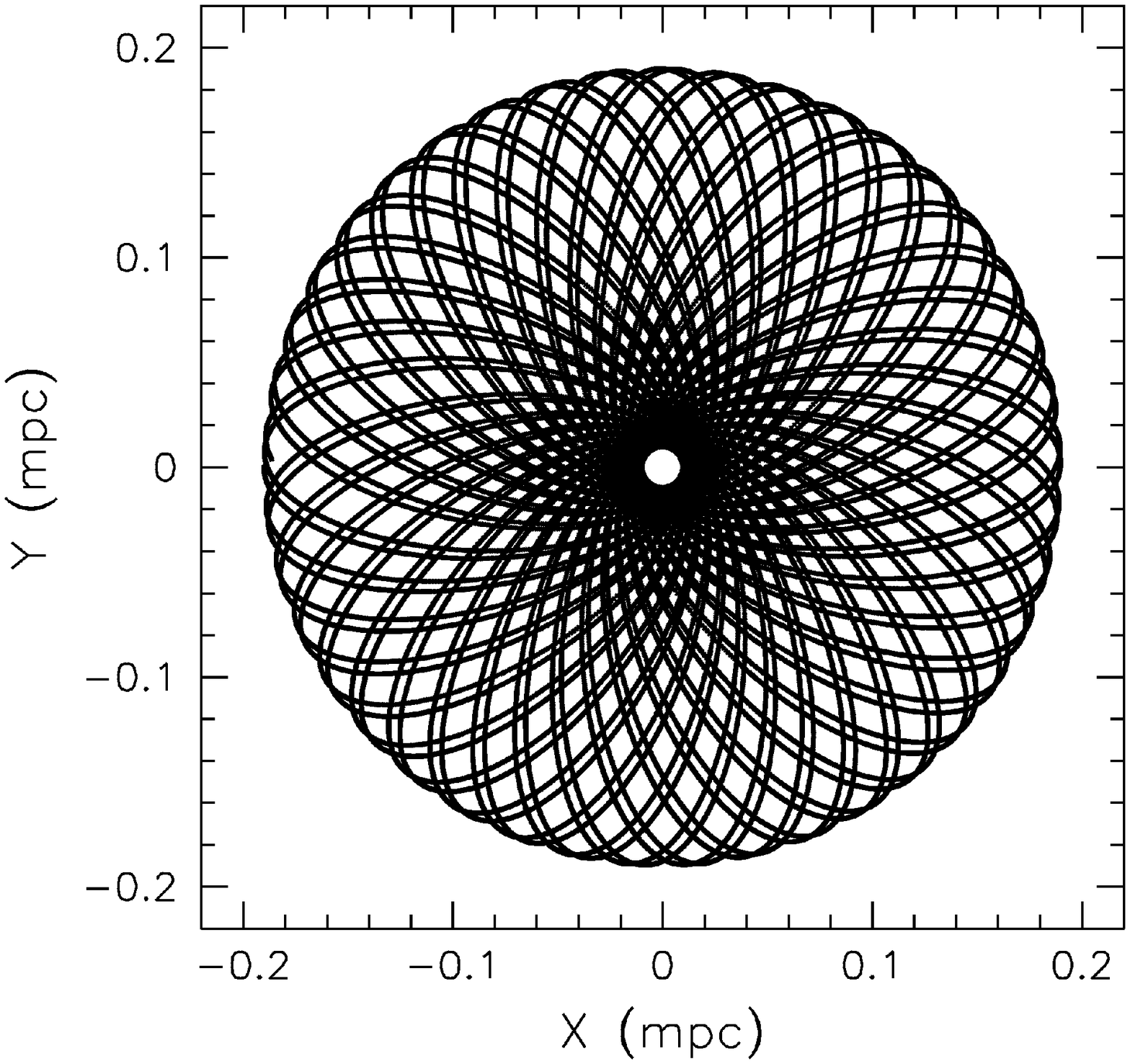}  
  \end{minipage}
  \begin{minipage}[b]{6 cm}
    \includegraphics[width=5cm]{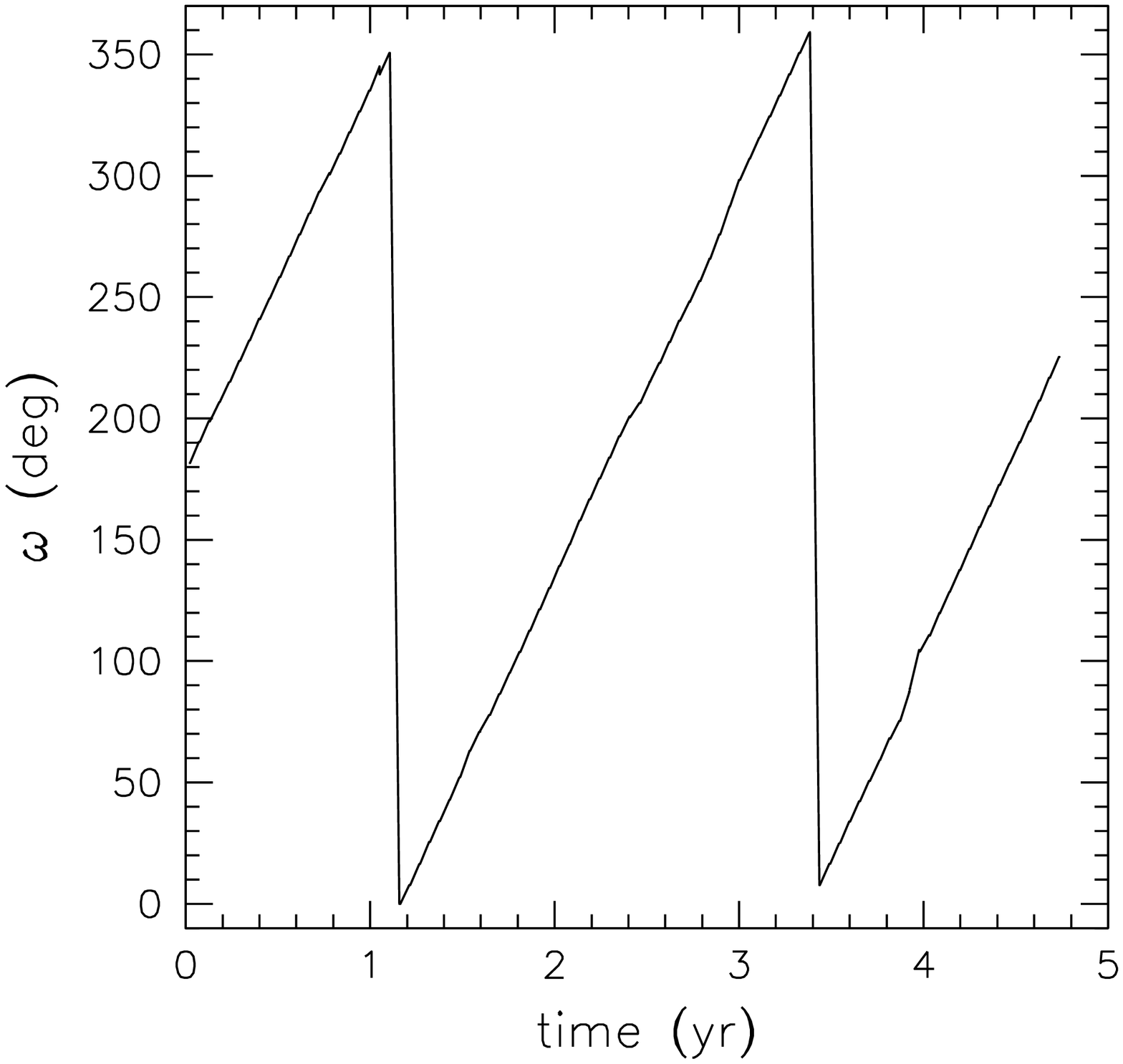}  
  \end{minipage}
  \caption{(Top) Trajectory of the IMBH over a time of 4.5 yr, showing
    precession of the periastron due to the PN terms.
(Bottom). Periastron advancement for the IMBH orbit. }
  \label{fig:traj}
\end{figure}

Figure~\ref{fig:dist} shows the time evolution of the distance 
between the IMBH and the SMBH in this integration, and in a 
second integration in which ARCHAIN was used to follow the binary in the
absence of stars.  It can be seen from Fig.~\ref{fig:dist} that
the stars have no significant effect on the rate of inspiral.
Figure~\ref{fig:traj} (top panel) shows the trajectory of the 
IMBH over a time of 4.5 yr, which roughly corresponds to the
time for $\omega$ to precess $360^\circ$ twice (bottom panel).
The longitude of the periastron is predicted to advance 
by an amount
\beq
\Delta \omega\approx 0.15^\circ \,(1-e^2)^{-1}
\left(\frac{\mbh}{3\times10^6\msun}\right)\left(\frac{\mpc}{a}\right)
\label{eq:advance}
\eeq 
each orbital period; this corresponds to $\sim 8^{\circ}$ per
revolution for $a=0.1\mpc$ and $e=0.9$.  

\begin{figure}
  \begin{center}
    \includegraphics[width=8cm]{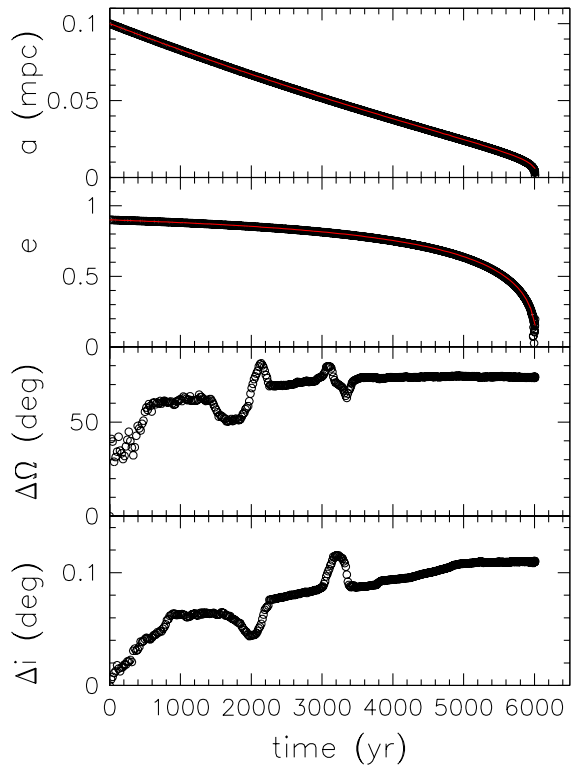}
  \end{center}
  \caption{Evolution of the Keplerian elements of the black hole binary 
    over the full inspiral time. The red lines in the upper two panels
    show the evolution of the semi-major axis and eccentricity 
    obtained by integration of Eqs.~\ref{eq:dadt} and \ref{eq:dedt}.}
  \label{fig:imbhk}
\end{figure}

In the absence of stars, evolution of the IMBH/SMBH binary would take
place in a fixed plane, i.e.  $i$ and $\Omega$ would remain constant.
In the presence of stars, however, deviations of the potential from
spherical symmetry cause the orientation of the binary to change with
time \citep{merritt-02}.  The evolution of the Keplerian elements of
the binary can be seen in Figure \ref{fig:imbhk}, which shows a
substantial change in the binary's orbital plane during the course of
the inspiral.  Similar evolution has been observed in other $N$-body
studies \citep[e.g.][]{MM-01,BGZ-06}.

Also shown in this figure are the evolution of the semi-major axis 
and eccentricity of the binary.
Results from $\varphi$GRAPEch are compared with 
numerical integrations of the coupled equations (Peters 1964)
\begin{equation}
\label{eq:dadt}
\frac{da}{dt} = -\frac{64}{5} \frac{G^3}{c^5} 
\frac{M_1\,M_2\,\left(M_1+M_2\right)}{a^3} F(e)
\end{equation}
\begin{equation}
\label{eq:dedt}
\frac{de}{dt} = -\frac{304}{15} \frac{G^3}{c^5} 
\frac{M_1\,M_2\,\left(M_1+M_2\right)}{a^3} G(e)
\end{equation}
where 
\begin{subequations}
\begin{eqnarray}
F(e) &=& \left(1-e^2\right)^{-7/2} \left(1+\frac{73}{24}e^2+
\frac{37}{96}e^4\right), \\
G(e) &=& \left(1-e^2\right)^{-5/2} \left(1+\frac{121}{304}e^2\right)\,.
\end{eqnarray}
\end{subequations}
The agreement is very good.

\begin{figure}
  \begin{center}
    \includegraphics[width=8cm]{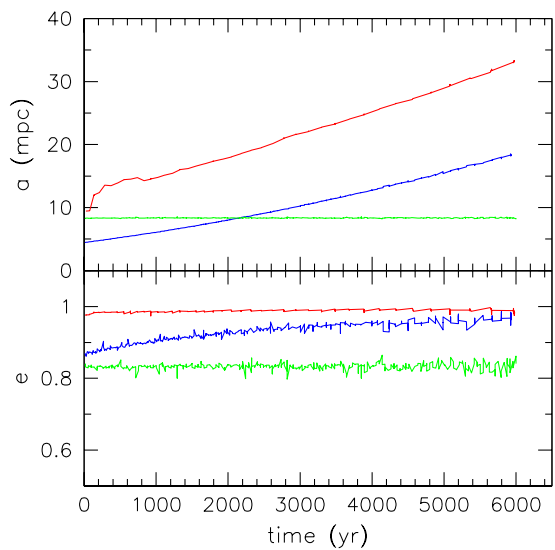}
  \end{center}
  \caption{Orbital elements of the S-stars S0-2 (blue), S0-16 (red)
    and S0-19 (green) over a time of 6000 yr.}
  \label{fig:sskpl}
\end{figure}

Figure~\ref{fig:sskpl} shows the evolution of the orbits of the three 
innermost S-stars in this simulation.
  The periapse distance of S0-16 is
$0.2\mpc$, roughly equal to the initial apoapse distance of the IMBH.
This orbit evolves strongly due to interactions with the IMBH.  The
semi-major axis increases by a factor of $\sim 4$ and the eccentricity
increases almost to one.  If the inspiral were prolonged, e.g. by
making the IMBH orbit less eccentric, Fig.~\ref{fig:sskpl} suggests
that this star and star S0-2 might be ejected completely after a few
tens of thousands of years \citep{MM-08}.

The binary ejects stars that interact strongly with it
and such ejections are a possible source of 
the so-called hyper-velocity stars \citep{Brown06}.
\begin{figure}
  \begin{center}
    \includegraphics[width=8cm]{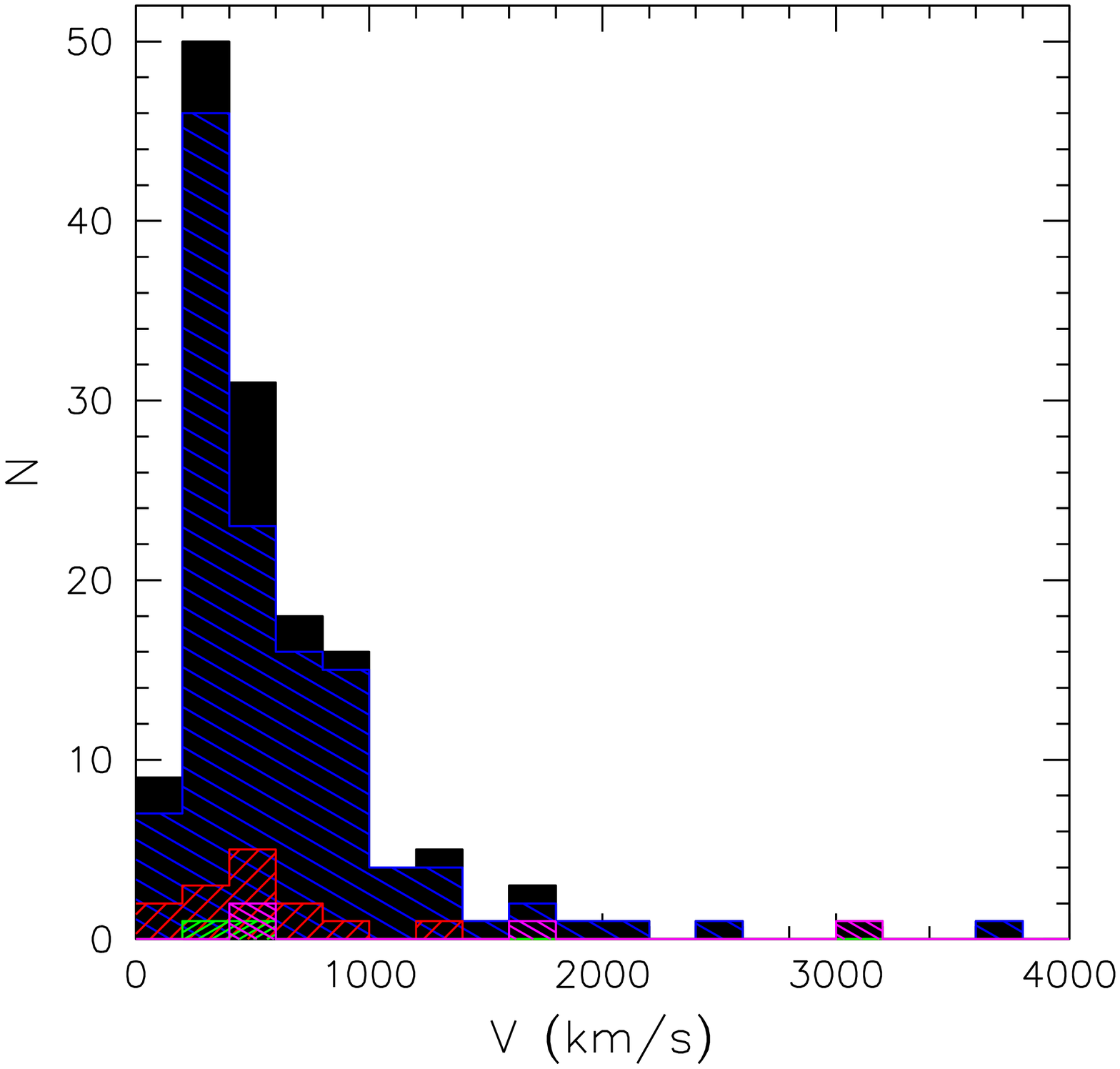}
  \end{center}
  \caption{Velocity distribution for stars and stellar remnants
    escaping from the SMBH at
    the end of the IMBH inspiral. The distribution for all escapers is
    presented in black, while the curves for individual species are:
    main-sequence (blue), white-dwarfs (red), neutron stars (green),
    black holes (magenta).}
  \label{fig:hvs}
\end{figure}
Figure~\ref{fig:hvs} shows the distribution of ejection velocities for
stars unbound to the SMBH at the end of the IMBH inspiral. 
The peak of the distribution is at $V_{\rm peak} \sim 300\kms$. 
Interestingly,
about 30\% of the ejected stars have velocities $\gap700\kms$,
i.e. large enough to escape the bulge and reach the Galactic halo as
hyper-velocity stars. 
About 150 stars are ejected during the inspiral,
which results in an average ejection rate of 
$\sim 20000\,\rm Myr^{-1}$.  
This rate is somewhat higher than observed in simulations that start
with a more separated binary \citep[e.g.][]{BGZ-06}, 
presumably because most of the
ejections we see are from stars on orbits that intersect the
binary at time zero, and many of these stars would have been
ejected at earlier times.
Most of the escapers are main-sequence stars, which is not surprising
given their dominance in the cluster.
Nonetheless, a handful of stellar
mass black holes are ejected with velocities up to $3000\kms$.

\section{Summary}

We have described a new $N$-body code, called $\varphi$GRAPEch,
designed for simulations of the central regions of galaxies containing
single or multiple supermassive black holes. 
Based on the serial implementation of $\varphi$GRAPE, the new code 
incorporates the algorithmic chain regularization scheme of \cite{MM-08}
to treat orbits near the central black hole(s) with
high precision. Post-Newtonian terms are included up to PN2.5 order. 

In performance tests, we find that the hybrid code
achieves better energy conservation in less computation time when
compared to the standard pure 4th-order Hermite integration scheme. A
simple performance model indicates that the computation time for
particles in the chain will dominate the total computation time if the
fraction of chain particles exceeds a certain limit. This limit is 
roughly 100 chain particles for a total number of particles of one
million.

We then apply our new hybrid code to a model of the Galactic center
that includes a supermassive black hole and four different stellar
mass components. The five shortest-period S-stars were also included
as test stars. We show that the orbits of S-stars are integrated with
very high precision. For the first time, we were able to measure the 
change in the orbital plane of S-stars due the effect of the finite-$N$ 
departures of the potential from spherical symmetry. The measured 
changes were in good agreement with theoretical predictions.

As a second application, we added an inspiraling intermediate-mass
black hole to our model of the Galactic center. The time scale of the
inspiral was found to be unaffected by the stars, in agreement with 
theoretical predictions given the small initial separation. 
However, the orbital plane of the inspiral was affected by perturabations
from the stars. A number of stars were ejected from the center 
with velocities large enough to reach the
Galactic halo as hyper-velocity stars.


{\bf Acknowledgments} The authors are very much indebted to S. Aarseth
and R. Spurzem for their many helpful suggestions about how to
incorporate AR-CHAIN in the $N$-body code.  We are also grateful to
T. Alexander for kindly providing the details of the galactic center
model used in \S 4. We thank the referee C.~Hopman for giving valuable
comments for improving the manuscript.  SH is supported by the NWO
Computational Science STARE project 643200503.  AG and DM were
supported by grants AST-0420920 and AST-0437519 from the NSF, grant
NNX07AH15G from NASA, and grant HST-AR-09519.01-A from STScI.

\bibliographystyle{mn2e}   
\bibliography{harfst}

\end{document}